\documentclass[preprint,tightenlines]{revtex4-1}
\usepackage{amssymb,amstext,appendix,hyperref,pgfplots,graphicx}
\usepackage{xcolor,subcaption}

\usepackage{amsmath}

\hypersetup{
colorlinks,
linkcolor=blue,
citecolor=blue,
urlcolor=blue}

\let\origleft\left
\let\origright\right

\renewcommand{\left}{\mathopen{}\mathclose\bgroup\origleft}
\renewcommand{\right}{\aftergroup\egroup\origright}

\newlength\figureheight
\newlength\figurewidth
\newlength\vertSkip
\newlength\horzSpace
\newcommand{\bv}{\mathbf{v}_{\mathbf{p}}}
\newcommand{\bvc}{\mathbf{v}_{\mathbf{p}^{c}}}
\newcommand{\bp}{\mathbf{p}}
\newcommand{\bE}{\mathbf{E}}
\newcommand{\rs}{\mathbf{r}_{s}^{(0)}}
\newcommand{\ta}{t_{a}}
\newcommand{\ts}{t_{s}}
\newcommand{\tsc}{t_{s}^{c}}
\newcommand{\tm}{t_{m}}
\newcommand{\tQ}{t_{Q}}
\newcommand{\tK}{t_{\kappa}}
\newcommand{\lr}{\mathrm{lr}}

\newcommand{\vu}{\hat{\mathbf{v}}_{\mathbf{p}}}
\newcommand{\dvC}{\Delta\mathbf{v}^{C}}
\newcommand{\bff}{\mathbf{f}}
\newcommand{\pL}{\left(}
\newcommand{\pR}{\right)}

\begin{document}
\title{Opportunities for detecting ring currents using the attoclock set-up}
\author{Jivesh Kaushal$^{1}$, Felipe Morales$^{1}$, and Olga Smirnova$^{1}$}
\address{$^{1}$Max Born Institute, Max Born Strasse 2a, 12489 Berlin, Germany}

\begin{abstract}
Strong field ionization by circularly polarized laser fields 
from initial states with internal orbital momentum has interesting propensity rule: electrons counter-rotating with respect to the laser field can be liberated more easily than co-rotating electrons [\href{http://dx.doi.org/10.1103/PhysRevA.84.063415}{Barth and Smirnova PRA {\bf 84}, 063415, 2011}]. 
Here we show that application of few-cycle IR pulses allows one to use this propensity rule to detect ring currents 
associated with such quantum states, by observing 
angular shifts of the ejected electrons. Such shifts present the main observable of the attoclock method. We use 
time-dependent Analytical $R$-Matrix (A$R$M) theory
to show that the attoclock measured angular shifts 
of an electron originating from two counter-rotating orbitals ($p^{+}$ and $p^{-}$) are noticeably different. Our work opens  new opportunities for detecting ring currents excited in atoms and molecules, using the attoclock set-up.
\end{abstract}
\maketitle

\section{Introduction}

Interaction of matter with strong, ultrashort laser fields offers new insights into phenomena that occur on the attosecond time-scale \cite{krausz2009}, providing new opportunities for probing electron dynamics \cite{schultze2010,pfeiffer2011}, structure of molecules \cite{spanner2004,meckel2008,blaga2012}, and deciphering the dynamics of laser-induced optical tunneling \cite{eckle2008,eckle2008-2,akagi2009, pfeiffer2011-2}.
The enticing opportunity to detect tunneling times during strong field ionization relies on attosecond angular streaking principle \cite{eckle2008,eckle2008-2,pfeiffer2012,pfeiffer2012}, which provides the link between electron detection angle and its time of ionization in strong infra-red circularly polarized fields. The application of ultra-short few-cycle pulses allows one to realize this principle experimentally in the so-called attoclock setup \cite{eckle2008, eckle2008-2, pfeiffer2012, pfeiffer2012}. 
Attoclock-based attosecond chronoscopy of strong field ionization can only be realized once the protocol for converting the attocklock observable -- the most probable electron detection angle -- into ionization time is clearly established \cite{torlina2015}. For a benchmark system, hydrogen atom, ionization time can be reconstructed using the combination of numerical and analytical approaches, leading to zero tunneling delays \cite{torlina2015}. However, such delays may become non-zero when several electrons are actively involved
in the ionization process \cite{torlina2015}, so that correlation-driven excitations during tunnel ionization \cite{torlina2012-2} are non-negligible.

New interesting questions that can be addressed by the attoclock include its sensitivity to internal electron dynamics prior to ionization. Perhaps the simplest example is the possibility of angular separation of photoelectrons originating from strong field ionization of two different orbitals carrying stationary currents. This is the case for, e.g., $p^{+}$ and $p^{-}$ orbitals in the ground states of noble gas atoms (Ar, Kr, Xe, etc.). Indeed, one would expect that when a co-rotating or a counter-rotating electron tunnel out, the rotating laser field will spun them 
away differently, as they have different initial velocities orthogonal to tunneling direction \cite{barth2011}. Interaction with the core potential should reveal this disparity in the initial conditions as it will affect the deflection angle, leading to angular separation of the photoelectron signals coming from the two orbitals. Here we demonstrate that this is indeed the case.

%

Since the seminal work of Keldysh \cite{keldysh1965}, the extended treatment by PPT \cite{ppt1966,ppt1967,pp1967,pkp1968} and subsequent theoretical efforts \cite{popru2008} to include the combined effects of the long-range core-potential and the strong laser field on the ionization process, several theoretical methods have been proposed to describe the nonadiabatic and nonlinear character of strong field ionization. Examples include the non-perturbative expansion techniques of Keldysh-Faisal-Reiss \cite{keldysh1965,faisal1973,reiss1980} and the Coulomb-Corrected Strong Field Approximation (CCSFA) method \cite{popru2008ii,popru2008iii,huismans2011}. The latter, in particular, relies on the imaginary time method (ITM) \cite{ppt1967} to develop a trajectory-based description of the ionization process including the Coulomb corrections to quantum trajectories. The CCSFA method, however, requires assumptions regarding initial conditions for these trajectories.

Here we develop further our Analytical $R$-Matrix (A$R$M) method \cite{torlina2012, torlina2012-2, kaushal2013, torlina2013}, a quantum-mechanical, gauge invariant approach which does not require any $a-priori$ assumptions regarding initial conditions for electron trajectories. The approach takes into account Coulomb interaction of the outgoing electron with the core within the time-dependent version of the Wentzel-Kramers-Brillouin (WKB) method. The concept of trajectories emerges naturally from the development of the theory and manifests in physically observable effects in the photoelectron spectrum \cite{torlina2013}. 

We present a theoretical description of attosecond angular streaking in long range potentials for orbitals carrying stationary current, extending our previous results \cite{torlina2015} beyond $s$-orbitals.
To explore theoretically the ionization from $p^{-}$ and $p^{+}$ orbitals in the attoclock setup, we extend our earlier long-pulse results \cite{kaushal2013}to the domain of short pulses. 
General theoretical analysis is complemented with additional results that include explicit expressions for momentum shifts due to the electron interaction with the Coulomb potential, for arbitrary final electron momentum, and detailed derivation of ionization delays in strong field ionization.

Crucially, for finite frequency of the ionizing circular field, the tunneling direction is not parallel to the laser field direction
at the moment of ionization. 
The associated 'tunneling angle' is determined by the direction of electron velocity at the complex-valued moment of time associated with the beginning of the tunneling process. This angle is also complex-valued.
Mathematically, 
unusual properties of strong field ionization from $p^{-}$ and $p^{+}$ states arise precisely from the contribution of the complex-valued 'tunneling angle' to ionization rates, as shown in \cite{barth2011} for short-range potentials. Notably, contribution
is absent for angle-independent $s$-states.
 
While the tunneling angle for $p^{-}$ and $p^{+}$ orbitals can be trivially found for short-range potentials \cite{barth2011}, the short-range model is unable to catch the key physics underlying the attoclock setup, manifest via the long-range electron-core interaction.
Within A$R$M, the key step in finding this angle is to establish the link between the final electron momentum $\mathbf{p}$ at the detector and the initial electron velocity that leads to this momentum $\mathbf{p}$, where the initial electron velocity is taken at the complex time associated with the beginning of the tunneling process. This link must include electron interaction with both the laser field and the Coulomb potential. We present a scheme that establishes such a link throughout the whole tunneling process, thus providing consistent treatment of long-range potential effects on photoelectron distributions from orbitals with arbitrary $(\ell,m)$ quantum numbers.

The knowledge of tunneling angle for long range potential requires the knowledge of distribution of initial momenta at the complex instant of time associated with the beginning of the tunneling process. Each point in this distribution is uniquely linked to the electron final momentum. In our previous work \cite{kaushal2013} we have established such link for the so-called \textit{optimal momentum}, which corresponds to the peak of the photoelectron distribution in long laser pulses. This was sufficient for describing long-pulse 
ionization dynamics discussed in \cite{kaushal2013}. 
However, for nearly single-cycle pulses and for states
with non-zero angular momentum this is 
no longer sufficient, even for the peak of the photoelectron spectrum, because such peak no longer corresponds to the ``optimal" momentum established for the 
nearly-monochromatic fields.
Thus, to obtain the attoclock spectra for $p^{-}$ 
and $p^{+}$ orbitals we need to refine our theory and 
establish the Coulomb corrections to the tunneling angle for every final momentum present in the attoclock spectrum.

The paper is organized as follows. Section~\ref{sec:key} describes key ideas that we have used to extend our A$R$M method to the case of short pulse ionization from the states with arbitrary $(\ell,m)$ in long-range potentials. Section~\ref{sec:results} describes our results. Appendices~\ref{app:theory} and \ref{app:dtsc} describe the key steps of our derivation. Particularly important are the derivations of the initial electron velocity and of 
the Coulomb correction to ionization time. Section~\ref{sec:conc} concludes the work.

\section{Key ideas of derivation} \label{sec:key}

Along the steps in deriving the long-pulse result in \cite{kaushal2013}, we encounter the following spatial integral [Eq.~(18) in \cite{kaushal2013}]:
\begin{multline}
a_{\mathrm{ARM}}(\mathbf{p},I_{P}) = \frac{i\kappa a^{2}}{(2\pi)^{3/2}}\int_{-\infty}^{T}dt'\int_{}^{}d\Omega'\,e^{-i\mathbf{v}_{\mathbf{p}}(t')\cdot\mathbf{a} - iS^{\mathrm{SFA}}(\mathbf{p},T;t')-iG_{C}(\mathbf{p},T;\mathbf{a},t')}\varphi_{\kappa\ell}(a) \times \\N_{\ell m}P_{\ell}^{m}(\cos\theta)e^{im\phi}. \label{eqn:amp_old_1}
\end{multline}
Here $\kappa=\sqrt{2I_p}$, $\varphi_{\kappa\ell}(a)$ is the radial wave-function at the $R$-matrix sphere of radius $r = a$ that separates the inner and outer $R$-matrix regions (see \cite{kaushal2013}), $N_{\ell m}$ is the spherical harmonic normalization coefficient, $S^{\mathrm{SFA}}$ is the well-known action in the Strong Field Approximation (SFA) \cite{keldysh1965,ppt1966,ppt1967,pp1967,pkp1968} for a free electron in a laser field and $G_{C}$ is the complex Coulomb phase correction, as introduced in \cite{kaushal2013,torlina2013}, $\bv(t)$ is electron velocity in the laser field. This result also holds for arbitrary field polarization and time profiles of the laser field envelope.
In Eq.~\eqref{eqn:amp_old_1} the integral is performed over solid angle of the sphere with radius $a$, where
the outgoing wavefunction outside the R-matrix sphere (in the so-called 'outer region') should match the wavefunction inside the 
R-matrix sphere (in the so-called 'inner region'). The boundary matching process has to ensure that the result is independent of the sphere radius $a$. In \cite{kaushal2013} this problem was solved for s-states, while for $p$-states it was only solved for an optimal momentum $\mathbf{p}_{\mathrm{opt}}$, which corresponds to the maximum of the photoelectron distribution in long pulses.

To find the matching scheme valid for any momentum $\mathbf{p}$, we write down the radial part of the asymptotic ground-state wavefunction
\begin{equation}
\varphi_{\kappa\ell}(r) = C_{\kappa\ell}\kappa^{3/2}e^{-\kappa r}(\kappa r)^{Q/\kappa - 1} = C_{\kappa\ell}\kappa^{3/2}e^{-iS_{C}(r)},
\end{equation}
where, $S_{C}(r) = -i(\kappa r - Q/\kappa\ln\kappa r) = S_{C}^{\mathrm{sr}} + S_{C}^{\mathrm{lr}}$, represents the complex, quantum-mechanical action derived through the Schr{\"o}dinger Equation in the asymptotic region $\kappa r \gg 1$, $C_{\kappa \ell}$ is the standard coefficient, determining the asymptotic behaviour of the radial wave-function. $S_{C}^{\mathrm{sr}}(r) = -i\kappa r$ is the short-range part of $S_{C}$, responsible for generating the complex momentum $\nabla S_{C}^{\mathrm{sr}} = -i\kappa\,\hat{\mathbf{r}}$ in the classically forbidden region.
Finally, $S_{C}^{\mathrm{lr}}$ provides the long-range potential correction to it, $\nabla S_{C}^{\mathrm{lr}}(r) = iQ/(\kappa r)\,\hat{\mathbf{r}}$, where $\hat{\mathbf{r}}$ is a unit vector of electron displacement. The main idea of wavefunction matching 
performed at the boundary $a$ of the R-matrix sphere 
(boundary matching) is to relate this long-range part of the action to the quasiclassical (WKB) action at 
every point on the $R$-matrix sphere, and at any time.

Along the lines of the derivation performed in \cite{kaushal2013}, we expand $G_{C}(\mathbf{p},T;\mathbf{a},t)$ and its long-range counterpart in $S_{C}(a)$ [$S_{C}^{\mathrm{lr}}(a)$ defined above], in a Taylor series around the SFA saddle points of integral Eq.~\eqref{eqn:amp_old_1}: $t_s'=t_a$, $\theta_{s}^{\prime}$, $\phi_{s}^{\prime}$ [see Eqs.~(20)-(22) of \cite{kaushal2013}]. As discussed in \cite{kaushal2013}, the saddle-point angles $\phi_s'(t')$, $\theta_s'(t')$ describe the direction of electron velocity at the time $t'$, and $t_a$ describes the time at which the electron trajectory crosses the $R$-matrix sphere boundary. Thus, there is a preferred direction along which the electron crosses the $R$-matrix sphere. This direction is given by the angles $\phi_s'(t')$, $\theta_s'(t')$, or simply by the vector:
\begin{equation}
\mathbf{r}_{s}^{(0)}(t)=a\hat{\mathbf{v}}_{\mathbf{p}}(t),
\label{eqn:rs}
\end{equation}
where $\vu(t)$ is a unit vector along the direction of electron velocity. This vector can be written in the equivalent form
\begin{equation}
\mathbf{r}_{s}^{(0)} = \int_{t_{s}}^{t_{a}}dt\,\mathbf{v}_{\mathbf{p}}(t),
\end{equation}
where $\ts$ is the SFA saddle point, which is a solution of equation $\partial S_{\mathrm{SFA}}/\partial t'=0$. The only non-negligible terms in a Taylor series around the spatial saddle points (as discussed in \cite{kaushal2013}), up to order $\mathcal{O}(a^{2})$ and $\mathcal{O}(Q^{2})$, are the following: 
\begin{multline}
S_{C}^{\mathrm{lr}}(a) + G_{C}(\mathbf{p},T;\mathbf{a},t) \simeq \left[S_{C}^{\mathrm{lr}}\left(\mathbf{r}_{s}^{(0)}\right) + G_{C}\left(\mathbf{p},T;\mathbf{r}_{s}^{(0)},t_a\right)\right] +\\ \left(\mathbf{a}-\mathbf{r}_{s}^{(0)}\right)\cdot \nabla \left.\left[S_{C}^{\mathrm{lr}}\left(\mathbf{r}\right) + G_{C}\left(\mathbf{p},T;\mathbf{r},t_{a}\right)\right]\right\vert_{\mathbf{r} = \mathbf{r}_{s}^{(0)}}. \label{eqn:ScGcTe}
\end{multline}

Boundary matching for the first group of terms has been performed in \cite{kaushal2013} and it simply yields:
\begin{equation}
W_{C}\pL \ts,\bp \pR \equiv \left[S_{C}^{\lr}\pL \rs \pR + G_{C}\left(\bp,T;\rs,\ta\right)\right] = \int_{t_{s}-i\kappa^{-2}}^{T}d\tau\,U\left(\int_{t_s}^{\tau}d\xi\,\bv(\xi)\right).
\label{eqn:W_C}
\end{equation}
The value of the lower limit of the integral $\tK = \ts - i/\kappa^{2}$ plays a key role in our ability to match the asymptotic ``tail" of bound wave-function $S_{C}^{\mathrm{lr}}(a)$ at the $R$-matrix sphere with the continuum ``tail" $G_{C}(\mathbf{p},T;\mathbf{a},t)$ and absorb them into one common 
expression, continuous across the matching boundary. This term represents the phase $W_C(t_s,\mathbf{p})$ of electron wave function accumulated as it travels from the atom to the detector.

Boundary matching for the momentum term requires a new approach, and is described in the Appendix~\ref{subapp:bmvel}. This approach generalizes the matching procedure for arbitrary order of terms 
in the Taylor series expansion Eq.~\eqref{eqn:ScGcTe}, and gives us closed-form expression for the Coulomb correction to 
the electron velocity $\bv \pL \ts \pR$ in a short-range 
potential:
\begin{equation}
-\dvC \equiv \left[\nabla S_{C}^{\mathrm{lr}}(\mathbf{r}_{s}^{(0)}) + \nabla G_{C}(\mathbf{p},T;\mathbf{r}_{s}^{(0)},t_a) \right] = \int_{t_s-iQ\kappa^{-3}}^{T}d\tau\,\nabla U\left(\int_{t_s}^{\tau}d\xi\,\mathbf{v}_{\mathbf{p}}(\xi)\right) + \bv\pL \ts \pR, \label{eqn:v_C}
\end{equation}
Note different value for the lower limit of the integral, $\tQ \equiv \ts - iQ/\kappa^{3}$. Equation~\eqref{eqn:v_C} represents the Coulomb correction to electron velocity $\bv(\ts)$, which takes into account the contributions of both the Coulomb and 
the laser fields at the complex time associated 
with the beginning of the tunneling process. 
We need to subtract $\dvC$ from the SFA velocity $\bv \pL \ts \pR$ to find the correction to initial velocity due to the long-range interactions, for a fixed momentum $\bp$ measured at the detector.

It might seem peculiar at first, as we first extend the domain of momentum generated by long-range potential deep under the barrier, and {\em then} subtract the short-range component in Eq.~\eqref{eqn:v_C}. The final velocity term contributing to the photoelectron angular distribution is $\bvc \pL \tsc \pR = \bv \pL \tsc \pR - \dvC$; this is to be contrasted with the case of the standard PPT \cite{ppt1966,ppt1967,pp1967,pkp1968} and KFR \cite{keldysh1965,faisal1973,reiss1980} theory, where the short-range SFA velocity $\bv(\ts)$ is the only source for the angular distributions and the prefactors in the ionization amplitudes/rates. 

The fact that the matching time for momentum, $t_{Q} = t_{s} - iQ/\kappa^{3}$, depends directly on the charge $Q$ in the zeroth order (unlike other complex times $\ta$, $\tK$ and $\ts$, which can only depend on $Q$ through higher-order Coulomb corrections, and not in the zeroth order), is a manifestation of the short-range contribution through a long-range potential expression in Eq.~\eqref{eqn:v_C}. In Appendix~\ref{app:SFAvp}, we further discuss this point, and show how $\dvC$ vanishes in the limit $Q \to 0$, which expresses the idea that physically, short- and long-range contributions are not separable effects, but need to be considered together to define the appropriate velocity generated in a Coulomb-laser coupled system.

The matching instant is different for the phase and its gradient (to wit, $t_{\kappa}$ and $t_{Q}$, respectively), which is not surprising, as different matching instants arise from different quantum boundary conditions: quasicalssical action, $G_{C}$, is matched to the action in the asymptotic limit, $S_{C}^{\mathrm{lr}}$, for the quantum mechanical wave function, whereas the gradient of the quasicalssical action, $\nabla G_{C}$, is matched to the gradient of the action stemming from the long-range part of the quantum mechanical wave function, $\nabla S_{C}^{\mathrm{lr}}$. A detailed derivation of boundary matching for the two cases is discussed in the Appendix~\ref{app:theory} and \ref{app:dtsc}.



After achieving the boundary matching for the momentum of the photoelectrons, we can resume the derivation scheme outlined in \cite{kaushal2013}, to end up with the final ionization amplitude:
\begin{equation}
a_{\mathrm{ARM}}(\mathbf{p},I_{P}) = (-1)^{m}C_{\kappa\ell}N_{\ell m}\sqrt{\frac{\kappa}{\left\vert S_{tt}\left(t_{s}^{c}\right) \right\vert}}e^{-iS^{\mathrm{SFA}} \pL \tsc,\bp \pR - iW_C \pL \tsc,\bp \pR} P_{\ell}^{m}\left(\frac{p_{z}^{c}}{v_{\mathbf{p}^{c}}\left(t_{s}^{c}\right)}\right)
e^{im\phi_{v}^{c}\left(t_{s}^{c}\right)}, \label{eqn:amp_new}
\end{equation}
which is applicable to any arbitrary final momentum $\mathbf{p}$, short pulses of arbitrary polarization, and initial states of arbitrary symmetry [arbitrary $(\ell,m)$ values], for a long-range interaction with the ionic core. The final Coulomb-corrected velocity entering into the argument of the prefactor $P_l^m$ 
and the tunnelling angle $\phi_{v}^{c}\left(t_{s}^{c}\right)$ is 
$\bvc \pL \tsc \pR = \bv \pL \tsc \pR - \dvC$. Finally, for
the coefficient  $\sqrt{\kappa}/\sqrt{|S_{tt}(t_s^c)}|$, 
where $S=S^{SFA}+G_C$,
we take its value in the tunnelling limit, where this
coefficient is
constant. (We also note that the leading term for 
$|S_{tt}(t_s^c)|$ is 
$|S_{tt}(t_s^c)|\simeq |\bv(\ts)\cdot\mathbf{E}(t_{s})|$ 
and since $|\bv(\ts)|=\kappa$,  this coefficient is 
indeed virtually constant near the field maximum.)

The Coulomb-corrected complex tunneling angle is: $\phi_{v}^{c}\left(t_{s}^{c}\right) \equiv \arctan[{v}^{y}_{\mathbf{p}^{c}}(t_{s}^{c})/{v}^x_{\mathbf{p}^{c}}(t_{s}^{c})]$. 	
The $c$ in the superscript of $t_{s}^{c}$ denotes the Coulomb-corrected saddle point $t_{s}^{c}\equiv t_{s}+\Delta t_s^{c}$ \cite{torlina2015}, which we derive in the Appendix~\ref{app:dtsc} and present another equivalent form for it:
\begin{equation}
\Delta t_s^{c} = -\left.\frac{d W_C(t_s,\mathbf{p})}{dI_p}\right\vert_{\kappa=\mathrm{const}} \equiv -\frac{\bv(\ts)\cdot\Delta\mathbf{v}^{C}}{\bv(\ts)\cdot\mathbf{E}(t_{s})}. \label{eqn:dtsc}
\end{equation}
The first equality in Eq.~\eqref{eqn:dtsc} has been used in \cite{torlina2015}, and independently derived through the proposd spin-orbit Larmor clock in \cite{kaushal2015}. The expression for $\dvC$, given by Eq.~\eqref{eqn:v_C}, is rigorously derived in Appendix~\ref{subapp:bmvel}, the essential point being that we can now describe the modifications of electron velocity due to long-range interactions {\em under the barrier} as the electron tunnels through.

\section{Results} \label{sec:results}

Fig.\textcolor{blue}{1 (a,b)} show the photoelectron spectra for strong-field ionization of $p^{-}/p^{+}$ orbitals of a Kr atom, i.e. for the Coulomb potential 
and the binding energy of Kr in Eq.~(\ref{eqn:amp_new}). 
We used right circularly polarized field, rotating
in the positive direction (counter-clockwise). The pulse was defined by its vector-potential ${\bf A}(t)$ as
\begin{equation}
\mathbf{A}(t) = -\frac{\mathcal{E}_{0}}{\omega}\cos^{2}\left( \frac{\omega t}{2N_{e}} \right) \left[\cos(\omega t)\,\hat{\mathbf{x}}+\sin(\omega t)\,\hat{\mathbf{y}} \right],
\end{equation}
with the envelope containing two full laser cycles base-to-base ($N_{e} = 2$) and the field envelope is modeled by a $cos^{2}$-profile.
The laser wavelength
was set to $\lambda=800$ nm, and the peak field
strength was set to $\mathcal{E}_{0} = 0.05$ a.u.
In these spectra, we can already identify several 
distinguishing features between ionization from the 
$p^{-}$ and $p^{+}$ orbitals. 

\begin{figure}
\centering
\includegraphics[width=1\textwidth]{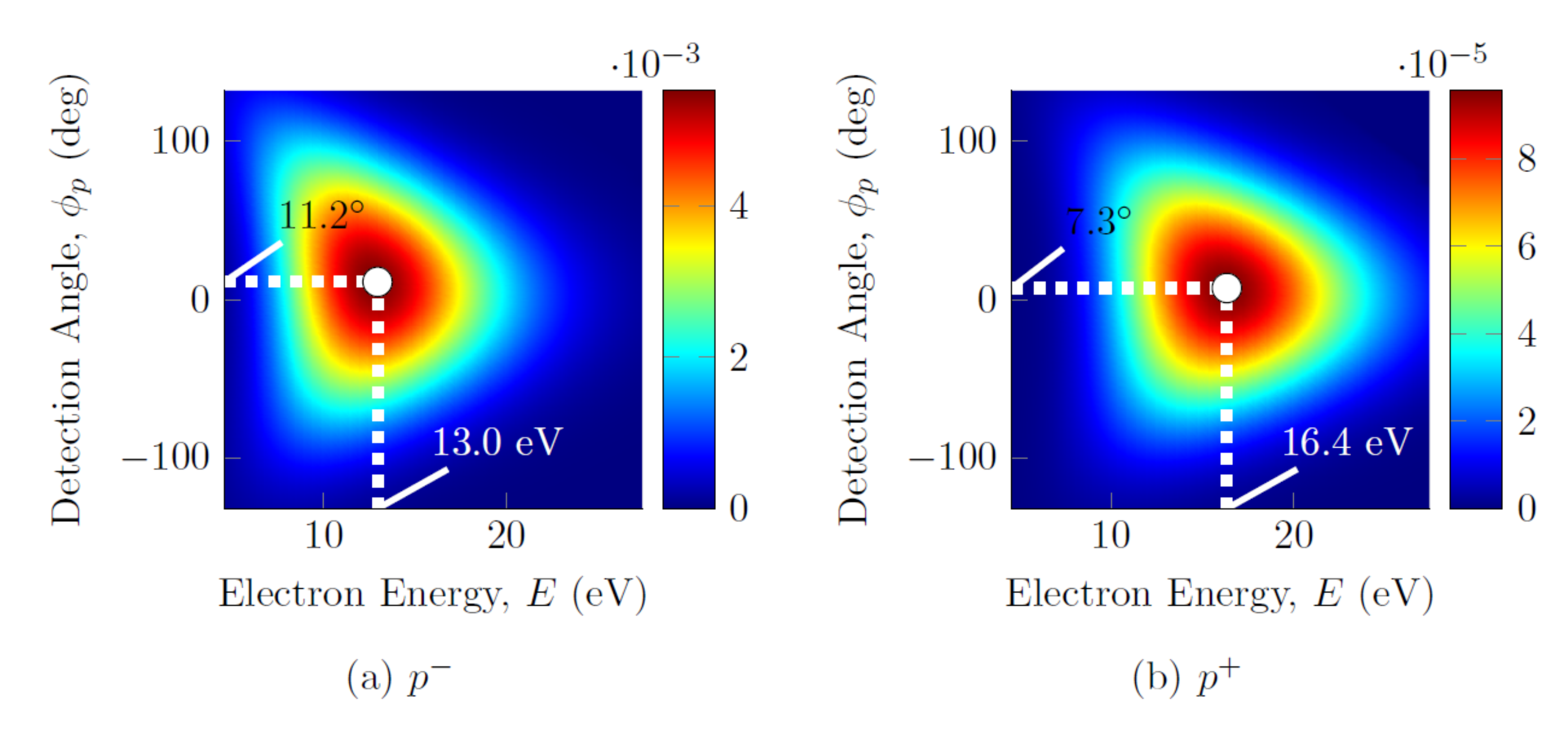}
\caption{Angle-resolved spectrum for $p^{-}$ and $p^{+}$ electrons removed from a Kr atom by a 2-cycle, right circularly-polarized field, for the peak intensity $I_{0} = 1.75 \times 10^{14}$ W/cm$^{2}$ ($\mathcal{E}_{0} = 0.05$ a.u.), $\lambda = 800$ nm } \label{fig:amap_1u0}
\end{figure}

Most important is the angular off-set between the peaks of photoelectron distributions corresponding to ionization from the $p^{-}$ and $p^{+}$ orbitals. Figure~\ref{fig:amap_1u0} shows that for the $p^{-}$ orbital, the spectrum is considerably offset, in the counter-clockwise direction, from the peak of the electron 
spectrum associated with the $p^{+}$ orbital. The difference between the offset angles is $\Delta\phi_{\mathrm{off}} \approx 3.9$ degrees, which is an observable shift. Note that ionization from short range potentials does not lead to angular off-set between these two spectra, where we would obtain the peak angle at $\phi_{\mathrm{off}} = 0$ degrees for both $p^{-}$ and $p^{+}$ orbitals.

This results can be understood as follows. Co-rotating and counter-rotating electrons are spun away by the attoclock differently, since they have different initial velocities orthogonal to tunneling direction. Interaction with the core potential reveals this disparity in the initial conditions: the slower counter-rotating electron stronger sinteract with the Coulomb field and is therefore deflected stronger. 
The co-rotating electron exits the barrier with 
higher lateral velocity, moves away faster, and is less affected by the Coulomb field. This leads to angular separation of the photoelectron signals from the two orbitals.

This picture is further confirmed by the shift of the peak energy for the two orbitals, which is also substantial, $\Delta E = 3.4$ eV, with $p^{+}$ electrons peaked at higher energy than $p^{-}$. For the present case of right-circularly polarized laser fields, the $p^{+}$ (or co-rotating, in general) electrons are always detected with higher energy than the $p^{-}$ (counter-rotating electrons) as discussed above. A similar energy off-set has been found in the case of long pulses and short-range potentials \cite{barth2011}. 

We can now explore the angular profile of photoelectron distributions in more detail, and study the distinguishing features of ionization from different orbitals. 
These features are determined by the long-range electron-core interaction.
The distribution for $p^{-}$ is more stretched out along the azimuthal angle $\phi_{p}$, but is more compact along the radial momentum $p_{\rho}$, as compared to the corresponding distribution in $p^{+}$. The spectrum spreads out in angle in case of $p^{-}$ orbital, because of the prominence of the low momentum electrons in that case, which leads to stronger Coulomb attraction and larger deflection angles. 

The width of the distribution in energy $E$ for the $p^{+}$ electron is greater than for the $p^{-}$ electron. 
This effect is related to the dominance of high energy electrons and specific energy dependence of the Coulomb effects. For high energies, the Coulomb effects weakly decrease with increasing energy, thus preserving the original Gaussian-like distribution for the $p^+$ spectra, characteristic of ionization
from short-range potentials. In contrast, photoelectrons emitted from $p^{-}$ orbitals ``pile up" at lower energies due to 
Coulomb attraction.

\begin{figure}
\includegraphics[width=0.5\textwidth]{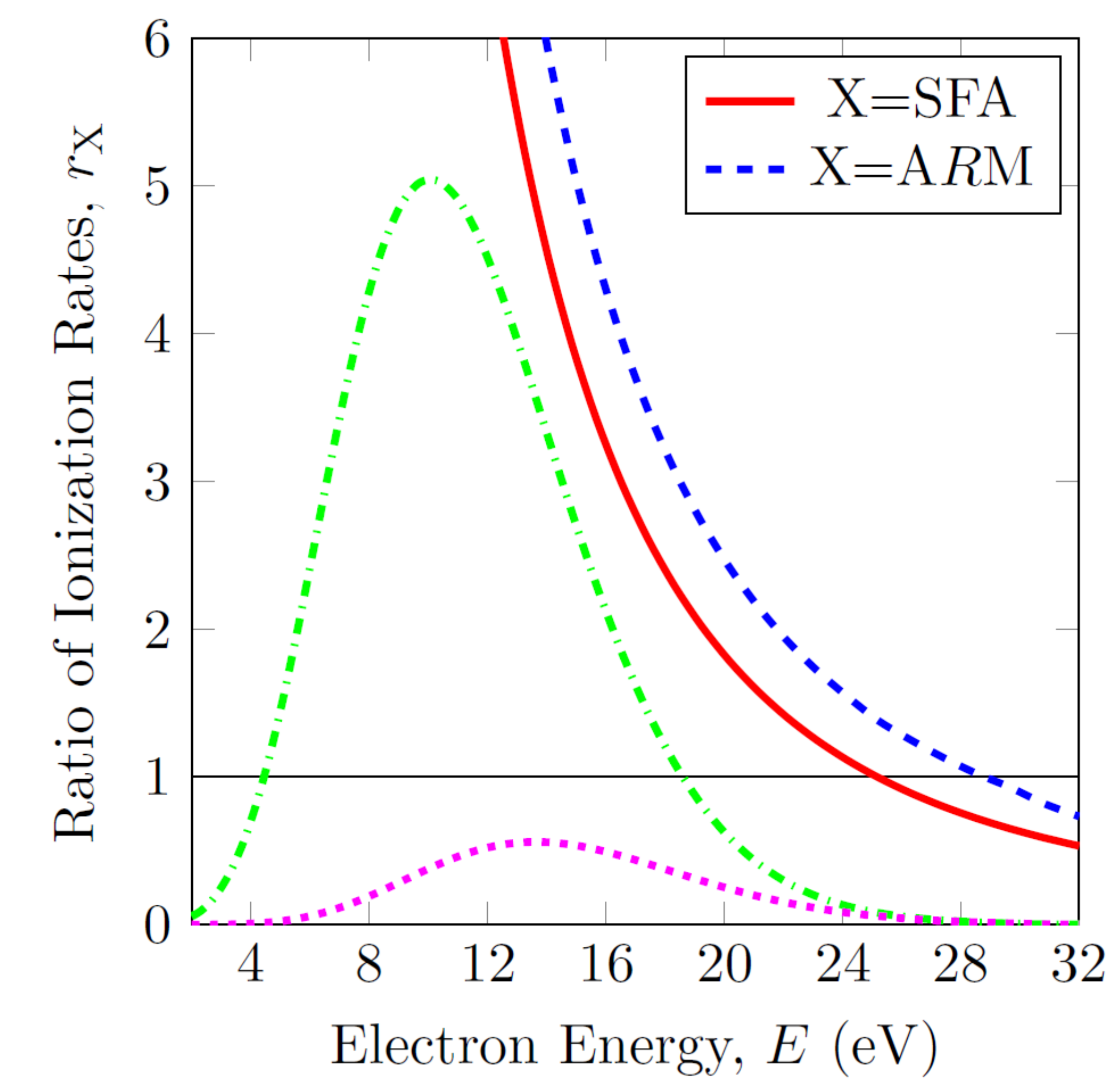}
\caption{Ratio of ionization rates for $p^{-}$ to $p^{+}$ orbitals, in SFA (red solid) and A$R$M blue dashed). The green dash-dotted curve is the photoelectron 
spectrum for the (dominant) $p^{-}$-orbital, the magenta dotted curve 
shows the spectrum for the $p^{+}$-orbital (in arbitrary units). The black thin solid line represents the unity 
level in the ratio. The calculations are for a 
Krypton atom at the peak intensity $I_{0} = 1.75 \times 10^{14}$ W/cm$^{2}$ ($\mathcal{E}_{0} = 0.05$ a.u.), $\lambda = 800$ nm, 2-cycle, right-circularly polarized laser field.} \label{fig:ratio_p-p+}
\end{figure}

A comparison of the ratio of angle-integrated ionization rates for $p^{-}$ and $p^{+}$ electrons, $r_{\mathrm{X}} = w_{\mathrm{X}}^{p^{-}}/w_{\mathrm{X}}^{p^{+}}$, is shown in Fig.~\ref{fig:ratio_p-p+}, for SFA (\textcolor{red}{red} solid) and A$R$M (\textcolor{blue}{blue} dashed). 

The green dashdotted curve is the photoelectron spectrum for the (dominant) $p^{-}$-orbital, in arbitrary units, and the black thin, solid line represents the unity level in the ratio. We observe considerable suppression of the signal from the $p^{+}$ orbital (nearly 2 to 6 times in the range from 25 to 15 eV, and more for lower energies), versus the signal from 
the $p^{-}$ orbital. Thus, when both orbitals are equally populated as is the case of the neutral Kr, the signal from
the $p^-$ orbital will dominate the total spectrum.

The dominance of the $p^-$ electron has the same origin as the one described in \cite{barth2011} for short-range potentials and long pulses. However, here the effect is further 
amplified due to the Coulomb effects. 
Low to medium energy photoelectrons are enhanced more 
strongly for $p^{-}$ than for $p^{+}$: for 16 eV 
electron energy, around 30 $\%$ more in 
the Coulomb potential compared to the short-range potential.
The region of dominance for $p^{-}$ orbital is 
slightly extended to higher energies, indicating the 
enhancement of nonadiabatic dynamics in the 
long-range potential. 
In \cite{barth2011}, it was expected that Coulomb corrections would nearly cancel out and thus have no impact on the ratio $w_{p^{-}}(E)/w_{p^{+}}(E)$, which we find to be true for higher energy photoelectrons, but not for low to medium energy.




The dominance of counter-rotating electrons over co-rotating ones implies that the former will give the prominent contributions to the ionization yields and angular distributions measured experimentally. 
In Fig.\textcolor{blue}{3 (a,b)} we show intensity scan of the offset angle and peak energy for 
Argon atom. The offset angles and peak energies for an $s$-orbital calculated with the same ionization potential of Argon would have
appeared approximately in the middle between the 
graphs presented in Fig.\textcolor{blue}{3 (a,b)}. 


\begin{figure}
\centering
\includegraphics[width=1\textwidth]{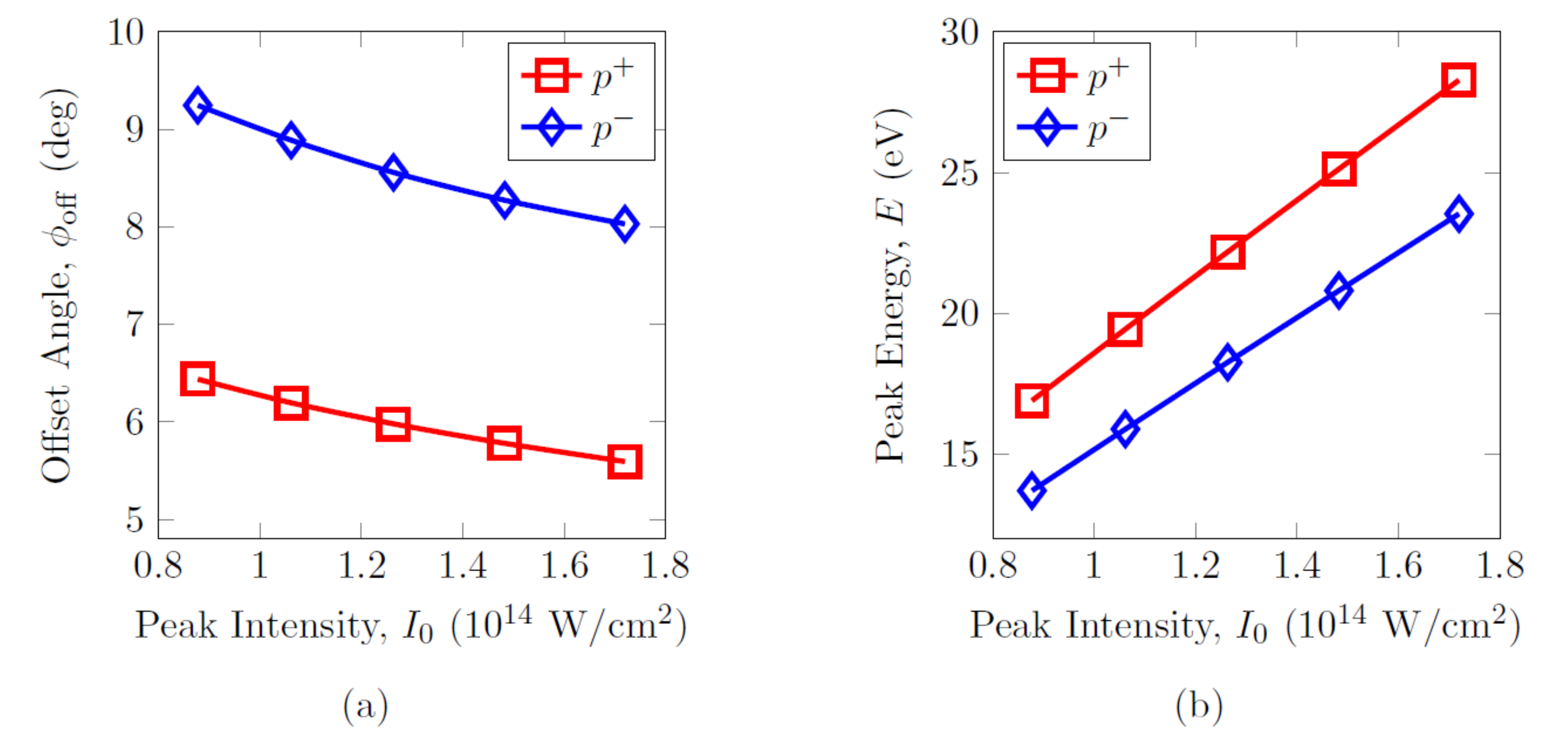}
\caption{(a) Offset Angle and (b) Peak energy variation for $p^{-}$ and $p^{+}$ electrons with peak intensity, for an Argon atom, $I_{P} = 15.76$ eV. Field Parameters: $\lambda = 800$ nm, 2-cycle, $\cos^{2}$-envelope, right-circularly polarized.}
\label{fig:phio_vs_I0}
\label{fig:Epeak_vs_I0}
\end{figure}

\section{Conclusions} \label{sec:conc}
We have extended our A$R$M method to include the effects of the 
long-range potential interactions on the outgoing electron with the core, for ionization from atomic orbitals of arbitrary symmetry,
going beyond $s$-orbital case considered in \cite{torlina2015}.
We have studied the effects of the initial 
orbital momentum on the observed final angle-and energy-resolved photo-electron distribution.

We have shown the sensitivity of the 
attoclock observables to the internal dynamics in the initial state. The difference between the attoclock off-set angles for $p^{+}$ and $p^{-}$ orbitals is about $3-4$ degrees for Ar and Kr in typical experimental conditions. 
Experimentally, the attoclock set-up has been applied to study ionization from $p$-states in Ar atom \cite{pfeiffer2012}. The resulting off-set angles have been used to extract the spatial coordinate corresponding to the position of the exit from the tunneling barrier. Our results suggest that
corrections at the level of about 2 degrees might be 
required for this mapping, since the off-set angle for an $s$-orbital
is about 2 degrees smaller than for the dominant 
$p^{-}$-orbital.


We expect that the attoclock set-up could be used for detecting ring currents, excited in atoms or molecules. Ring currents of opposite direction are expected to increase or decrease the attoclock off-set angle relative to the value detected in a system, in which such currents have not been excited in the initial state.

The direction of the stationary current in the initial state is also mapped onto the strength of the signal, with the signal from a current counter-rotating with respect to the laser field dominating over the co-rotating one. 

To increase the sensitivity of detecting the current direction, 
one can also measure angular and energy dependent photoelectron dichroism. It amounts to detecting the attoclock spectra in 
left and right circularly polarized fields and taking the ratio of the difference to the sum of such spectra. 
The resulting CD attoclock spectra will have opposite off-set angles for opposite directions of ring currents.



Finally, we note that energy separation of $p^{+}$ and $p^{-}$ signals in long pulses leads to spin-polarization \cite{barth2013}, thus angular separation should lead to additional opportunities to create short spin-polarized electron bunches. 

\section*{Acknowledgements}
We thank Lisa Torlina and Misha Ivanov for useful discussions.
J.K. and O.S. acknowledge support of the DFG project SM 292/2-3.
F. M. and O. S. acknowledge support of the DFG project SM 292/3-1.

\appendix

\section{Boundary matching for the gradient of EVA phase}
\label{app:theory}

\subsection{Initial velocity and tunneling angle} \label{subapp:bmvel}

We first derive Eq.~(\ref{eqn:v_C}). Since $\nabla G_{C}\pL\rs,\ta\pR$ is to be matched to $\nabla S_{C}^{\lr}\pL\rs,\ta\pR$, we have:
\begin{equation}
\nabla S_{C}^{\lr}\pL \rs,\ta \pR + \nabla G_{C}\pL \rs,\ta \pR = i\frac{Q}{\kappa a}\vu(\ts) + \int_{t_{a}}^{T} d\tau\,\nabla U\pL \int_{t_{s}}^{\tau}d\xi\,\mathbf{v}(\xi) \pR. \label{eqn:ScGcT}
\end{equation}
We have used that $\nabla S_{C}^{\mathrm{sr}} = -i\kappa\,\hat{\mathbf{r}}$, and $\rs$ is given by Eq.~(\ref{eqn:rs}).
 The change of the unit vector of velocity $\vu(\ta)$ in $\nabla S_{C}^{\lr}$  to $\vu(\ts)$ is validated by the fact that the first order term proportional to $(\ta - \ts)$ has exactly zero contribution to the matching point, regardless of the duration and envelope profile of the field. The second order corrections are $\propto \mathbf{E}^{2}$, effects which we exclude as they require consideration of the polarization of the bound state \cite{murray2010}. We require
\begin{equation}
\nabla S_{C}^{\mathrm{lr}}\pL \rs,\ta \pR = \int_{t_{m}}^{t_a}d\tau\,\nabla U\left(\int_{t_s}^{\tau}d\xi\,\mathbf{v}_{\mathbf{p}}\left(\xi\right)\right) + \mathbf{f}, \label{SU}
\end{equation}
where $t_{m}$ is an unknown complex matching time instant that we have to establish. Once we derive the expression for $t_{m}$, we can combine $\nabla S_{C}\pL \rs,\ta \pR+\nabla G_{C}\pL \bp,T;\rs,\ta \pR$ into a single term:
\begin{equation}
\nabla S_{C}\pL \rs,\ta \pR + \nabla G_{C}\pL \mathbf{p},T;\rs,\ta \pR = \int_{t_{m}}^{T}d\tau\,\nabla U\pL \int_{t_{s}}^{\tau}d\zeta\,\bv \pL \zeta \pR \pR + \mathbf{f}. \label{result}
\end{equation}

Here we have made allowance for an additional, constant vector $\bff$ that will aid us in our matching scheme. 
The idea of the matching scheme is to redistribute the contributions from the terms appearing in the RHS of Eq.~\eqref{eqn:ScGcT} in a boundary-independent form, to which purpose the constant vector $\bff$ is introduced. The choice of $\bff$ depends on the choice of the matching instant $\tm$, which as will be shown, we are free to decide upon; however, a specific choice of $\tm$ leads to a clear physical interpretation, and hence is favoured.

First, we note that we can rewrite the integral on the 
RHS of Eq.~\ref{SU} using the short-time approximation for the argument of $U(\mathbf{r})=-Q/\left\|r\right\|$, which is justified since time instants $t_s$ and $t_a$ are very close to each other by construction: $\vert \ts - \ta \vert = a/\kappa \ll \vert \ts \vert$. For any $\tau$ between $t_{s}$ and $t_{a}$ this approximation yields:
\begin{equation}
\left\|\int_{t_s}^{\tau}d\zeta\,\mathbf{v}_{\mathbf{p}}(\zeta)\right\|\approx\|\mathbf{v}_{\mathbf{p}}(t_s)\|\left(\tau-t_s\right) = i\kappa\left(\tau-t_s\right) = \int_{t_s}^{\tau}d\zeta\,i\kappa, \label{approx}
\end{equation}
using $\|\bv(\ts)\|\equiv v_{\mathbf{p}}=i\kappa$. From Eq.~\eqref{approx} we obtain:
\begin{equation}
\int_{\tm}^{\ta}d\tau\,\nabla U\left(\int_{\ts}^{\tau}d\xi\,\mathbf{v}_{\mathbf{p}}\left(\xi\right)\right) = Q\frac{\mathbf{v}_{\mathbf{p}}(\ts)}{v_{\mathbf{p}}^{3}(\ts)}\int_{\tm}^{t_{a}}dt'\,\frac{1}{(t' - \ts)^{2}}, \label{Uexpl}
\end{equation}
	
We therefore obtain the condition for the matching point $\tm$, using Eqs.~(\ref{eqn:ScGcT}) and (\ref{Uexpl}) to rewrite Eqs.~(\ref{SU}) as:
\begin{equation}
i\frac{Q}{\kappa a}\frac{\mathbf{v}(t_{s})}{v(t_{s})} = -Q\frac{\mathbf{v}(t_{s})}{v^{3}(t_{s})}\left[\frac{1}{t_{a}-t_{s}} - \frac{1}{\tm - t_{s}}\right] + \bff.
\end{equation}
Using the definition of $t_{a} = t_{s} - ia/\kappa$, the first term on the RHS cancels with the expression on the 
LHS, giving us the following definition for the 
time instant $t_{m}$:
\begin{equation}
\bff = \frac{Q}{\kappa^{2}}\frac{1}{\tm - \ts}\vu(\ts) \Rightarrow \tm = \ts + \frac{Q}{\kappa^{2} \bff\cdot\vu(\ts)}.
\end{equation}
The first and most obvious choice of a suitable vector $\bff$ we can consider is what we see time and again in strong-field ionization: $\bff = \bv(\ts)$ (the SFA velocity), which gives us
\begin{equation}
\tm = \tQ \equiv \ts - i\frac{Q}{\kappa^{3}}.
\end{equation}
With this definition of matching point, the boundary-independent momentum contribution from long-range part is also clearly stated:
\begin{equation}
-\dvC \equiv \nabla S_{C}^{\lr} + \nabla G_{C} = \int_{\tQ}^{T}d\tau\,\nabla U\left(\int_{\ts}^{\tau}d\xi\,\bv(\xi)\right) + \bv(\ts), \label{eqn:vC}
\end{equation}
\begin{equation}
\Delta \mathbf{v}^{C} = \bv^{C} \pL \ts \pR - \bv \pL \ts \pR,
\end{equation}
\begin{equation}
\bv^{C} \pL \ts \pR=-\int_{\tQ}^{T}d\tau\,\nabla U\left(\int_{\ts}^{\tau}d\xi\,\bv(\xi)\right).
\end{equation}
$\bv^{C} \pL \ts \pR$ describes the electron velocity that includes coupled contributions of "laser-free" bound velocity and "laser-driven" continuum velocity. 

The part associated with the contribution from the long-range potential obtains by subtracting the SFA velocity $\bv \pL \ts \pR$ from $\bv^{C} \pL \ts \pR$.  We have also defined, along with our matching scheme, a clear definition of momentum shifts induced by long-range interaction, and, the crucial point of all, taking into account the contributions from under the barrier motion 
to the momentum shifts induced by the Coulomb potential.

Substituting this value of $\tm$ into Eq.~\eqref{result} we obtain Eq.~\eqref{eqn:v_C}. Tunneling angle is obtained from Eq.~\eqref{eqn:v_C} as described in the main text.


\subsection{Obtaining the SFA velocity from $\bv^{C} \pL \ts \pR$ in the limit $Q \to 0$} \label{app:SFAvp}

We underscore the peculiarity of the matching time $t_{Q}$: it is the only complex time discussed here that explicitly depends on the charge $Q$ in zeroth order; $t_{a}$, $t_{\kappa}$ and $t_{s}$ are all independent of the effective long-range charge. The expression Eq.~\eqref{eqn:v_C} not only contains the long-, but also the short-range contribution, which is the source of the complex velocity $\mathbf{v_{p}}(t_{s}) = i\kappa\,\hat{\mathbf{v}}_{\mathbf{p}}(t)$. Therefore, in the limit of a short-range potential ($Q \to 0$), the long-range contribution in Eq.~\eqref{eqn:v_C} should converge to zero.

To demonstrate that this is indeed the case, we divide the integral in Eq.~\eqref{eqn:v_C} into two parts: 
the integral from $t_{Q}$ till the matching point $t_{a}$, up to which time we use the asymptotic, quantum action $S_{C}$ for the wavefunction, and beyond which the quasiclassical action is used leading to the eikonal-Volkov \citep{smirnova2008} phase contribution. With $Q \to 0$ this latter part converges to zero as it is directly proportional to $Q$. From the former, we get:
\begin{equation}
\lim_{Q \to 0} \mathbf{v}_{\mathbf{p}}^{C} = -\lim_{Q \to 0} \int_{t_{Q}}^{t_{a}}d\tau\,\nabla U(\mathbf{r}_{L}) \approx \lim_{Q \to 0} Q\frac{\mathbf{v_{p}}(t_{s})}{v_{\mathbf{p}}^{3}(t_{s})}\left[\frac{1}{t_{a} - t_{s}} - \frac{1}{t_{Q} - t_{s}}\right] \label{eqn:vpC_exp}
\end{equation}
after approximating the trajectory by its first (linear) order term in time, on account of the proximity of $t_{Q}$ and $t_{a}$ in the complex-time plane.

In Eq.~\eqref{eqn:vpC_exp}, the first term goes to zero, since there is no dependence on charge $Q$ in $t_{a}$ or $t_{s}$. If we were considering higher order corrections to $t_{a}$, $t_{s}$, even then the first term in Eq.~\eqref{eqn:vpC_exp} would converge to zero, since the zeroth order term (independent of $Q$) will prevail in that case over the higher order correction (dependent on $Q$), leading to finite contribution from the first term even when $Q \to 0$.

The same is not true for the second term in the rectangular brackets of Eq.~\eqref{eqn:vpC_exp}, because of $t_{Q} = t_{s} - iQ\kappa^{-3}$, and using this definition of $t_{Q}$, we get:
\begin{equation}
\lim_{Q \to 0}\mathbf{v}_{\mathbf{p}}^{C}= -\lim_{Q \to 0} Q\frac{\mathbf{v_{p}}(t_{s})}{v_{\mathbf{p}}^{3}(t_{s})}\frac{i\kappa^{3}}{Q} = \mathbf{v_{p}}(t_{s})
\end{equation}
which is the SFA velocity, and is precisely what we have 
intended to prove in the limit of $Q \to 0$. This term then cancels with $\bv \pL \ts \pR$ in Eq.~\eqref{eqn:v_C} to give $\dvC = 0$ in the short-range limit.

\section{Derivation of Coulomb correction to ionization time} \label{app:dtsc}

We start with Eq.~(35) of \cite{kaushal2013} for the Coulomb correction to the saddle-point time $t_{a}$, corresponding to the moment of time when trajectory crosses the boundary of the R-matrix region:
\begin{equation}
\Delta \ta^{c} = -\frac{\partial_{t}G_{C}\pL \rs,\ta \pR}{\partial_{t}^{2} S^{\mathrm{SFA}}\pL \rs,\ta \pR} = \frac{-\mathbf{v}_{\mathbf{p}}\left(t_s\right)\cdot\dvC\left(t_a,T\right) + U(a)}{\mathbf{E}\left(t_s\right)\cdot\mathbf{v}_{\mathbf{p}}\left(t_s\right)} \label{dt_anew}
\end{equation}
We can rewrite this equation as
\begin{equation}
\Delta \ta^{c} = \frac{\bv\pL \ts \pR \cdot \int_{\ta}^{T}d\tau\,\nabla U\pL \int_{\ts}^{\tau}d\xi\,\bv(\xi) \pR + U(a)}{\mathbf{E}\pL \ts \pR\cdot\bv\pL \ts \pR}. \label{dt_anew01}
\end{equation}
The following is true for any $\tm$ between $\ts$ and $\ta$:	
\begin{equation}
\bv \pL \tm \pR\cdot\int_{\tm}^{\ta}dt \nabla U\pL \int_{\tm}^{t}d\tau \bv \pL \tau \pR \pR \simeq \frac{Q}{\bv(\ts)}\left[\frac{1}{\ta - \ts} - \frac{1}{\tm - \ts}\right]=\frac{Q}{i\kappa(\tm - \ts)} - \frac{Q}{a}, \label{connection}
\end{equation}
on account of the short-time approximation as outlined above, see Eq.~\eqref{approx}. Taking into account that Eq.~(\ref{connection}) yields:
\begin{equation}
U(a)=- \frac{Q}{a}=\bv \pL \tm \pR\cdot\int_{\tm}^{\ta}dt \nabla U\pL \int_{\tm}^{t}d\tau \bv \pL \tau \pR \pR -\frac{Q}{i\kappa(\tm - \ts)} , \label{connection}
\end{equation}
we can rewrite Eq.~(\ref{dt_anew01}) as:
\begin{equation}
\Delta \ta^{c} = \frac{\bv\pL \ts \pR\cdot\int_{\tm}^{T}dt \nabla U \pL\int_{t_s}^{t}d\tau \bv\pL \tau \pR \pR + U \pL \int_{\ts}^{\tm}d\tau \bv \pL \tau \pR \pR}{\mathbf{E}\pL \ts \pR\cdot \bv \pL \ts \pR}, \label{dt_anew1}
\end{equation}
Here $\tm$ denotes any arbitrary complex moment of time that has to be established. Note that now the Coulomb correction $\Delta t_a$ to saddle point time $t_a$ given by Eq.~\eqref{dt_anew1} does not depend on the position of the boundary, but depends on 
time $\tm$.

We now have a similar freedom in choosing $\tm$ for the phase, as we had for the momentum. At present, we consider the equivalent expressions obtained from two different choices of $\tm$.

Taking $\tm = \tK$, the time instant for phase matching (as derived in Appendix~\ref{subapp:bmGc}), we get:
\begin{equation}
\Delta t_{s}^{c} = \frac{\mathbf{v}_{\mathbf{p}}\left(t_s\right)\cdot\int_{t_{\kappa}}^{T}dt\,\nabla U(\int_{t_s}^{t}d\tau \mathbf{v}_{\mathbf{p}}(\tau)) + U(\int_{t_s}^{t_{\kappa}}d\tau \mathbf{v}_{\mathbf{p}}(\tau))}{\mathbf{E}\left(t_s\right)\cdot\mathbf{v}_{\mathbf{p}}\left(t_s\right)} \label{dt_anew3}
\end{equation}
Note that Eq.~\eqref{dt_anew3} can be written in a compact form: 
\begin{equation}
\Delta t_s^{c} = -\left.\frac{d G_{C}(\mathbf{p},T,\mathbf{r}_{s},t_{\kappa})}{dI_p}\right\vert_{\kappa=\mathrm{const}}. \label{dt_anew4}
\end{equation}
This form has been applied in \cite{torlina2015}, where we have used notation
\begin{equation}
W_C(t_s,\mathbf{p})\equiv G_{C}\left(\mathbf{p},T,\mathbf{r}_{s}^{(0)},t_{\kappa}\right) = \int_{t_s-i/\kappa^2}^{T}d\tau U\left(\int_{t_s}^{\tau}d\zeta\,\mathbf{v}_{\mathbf{p}}\left(\zeta\right)\right).
\end{equation}
The time instant $\tK$ is when the electron is at a distance of $1/\kappa$ from the entrance point of the tunneling barrier.

With another choice of $\tm = \tQ$, we however get:
\begin{equation}
\Delta \ts^{c} = \frac{\bv\pL \ts \pR \cdot \int_{\tQ}^{T}dt\,\nabla U\pL \int_{\ts}^{t}d\tau\,\bv\pL \tau \pR \pR + U\pL \int_{\ts}^{\tQ}d\tau\,\bv\pL \tau \pR \pR}{\bE\pL \ts \pR \cdot \bv\pL \ts \pR} \label{eqn:dts2}
\end{equation}
Using the fact that $U\pL \int_{\ts}^{\tQ}dt\,\bv(t) \pR \approx i\kappa \vu$, we can rewrite Eq.~\eqref{eqn:dts2} in an equivalent form:
\begin{equation}
\Delta \ts^{c} = -\frac{\bv\pL \ts \pR \cdot \dvC}{\bv\pL \ts \pR \cdot \bE\pL \ts \pR}.
\end{equation}
We have used here the definition of $\dvC$ derived in Appendix~\ref{subapp:bmvel}. A similar result was derived for the optimal momentum in \cite{kaushal2013}, and now with the matching scheme presented here (which is a general procedure for matching to arbitrary orders of atomic charge $Q$), we have a rigorous derivation valid for arbitrary final momentum $\bp$ at the detector. We stress that for the hydrogen atom, where 
$Q=1$ and $\kappa=1$, both expressions are equivalent, since
$\tQ = \tK$.


\subsection{Boundary matching for the Coulomb phase} \label{subapp:bmGc}

Here we reproduce the boundary matching for the Coulomb phase term $G_{C}\pL \bp,T;\rs,\ta \pR$, as done in \cite{kaushal2013}, and present a generalization for the matching scheme for the 
phase as well.
 
The idea of matching for the phase is very similar to the idea of matching for its gradient discussed above. To emphasise this similarity we will use the same steps in our derivation, and show the flexibility in choosing the matching time instant $\tm$ for the phase, as we did for momentum.
 
Recalling that
\begin{equation}
G_{C}\left(\mathbf{p},T;\mathbf{r}_{s}^{(0)},t_a\right) = \int_{t_a}^{T}d\tau\, U\left(\mathbf{r}_{L}\left(\tau,\mathbf{r}_{s}^{(0)},\mathbf{p},t_a\right)\right),
\end{equation}
where
\begin{equation}
\mathbf{r}_{L}\left(\tau,\mathbf{r}_{s}^{(0)},\mathbf{p},t_a\right) =\mathbf{r}_{s}^{(0)} + \int_{t_a}^{\tau}d\xi\,\mathbf{v}_{\mathbf{p}}\left(\xi\right)=\int_{t_s}^{\tau}d\xi\,\mathbf{v}_{\mathbf{p}}\left(\xi\right),
\end{equation}
and $\mathbf{r}_{s}^{(0)}=\int_{t_s}^{t_a}d\tau\,\mathbf{v}_{\mathbf{p}}\left(\tau\right)$, we require that
\begin{equation}
S_{C}^{\lr}\pL \rs \pR = \int_{\tm}^{t_a}d\tau\,U\left(\int_{t_s}^{\tau}d\xi\,\mathbf{v}_{\mathbf{p}}\left(\xi\right)\right) + \Phi, \label{phSU}
\end{equation}
where $\tm$ is an unknown complex matching time instant, that we have to establish by fixing $\Phi$ along with it. Once it is found, we can combine $S_{C}^{\lr}\pL \rs,\ta \pR + G_{C}\pL \bp,T;\rs,\ta \pR$ into a single term:
\begin{equation}
S_{C}^{\lr}\left(\mathbf{r}_{s}^{(0)}\right) + G_{C}\left(\mathbf{p},T;\mathbf{r}_{s}^{(0)},t_a\right) = \int_{\tm}^{T}d\tau\, U\left(\int_{t_s}^{\tau}d\xi\,\mathbf{v}_{\mathbf{p}}\left(\xi\right)\right) + \Phi. \label{phresult}
\end{equation}
Expanding the RHS of Eq.~\eqref{phSU} using the proximity of $\ta$ and $\tm$ to $\ts$, we get
\begin{equation}
i\frac{Q}{\kappa}\ln(\kappa a) = i\frac{Q}{\kappa}\ln\pL \frac{a}{\kappa \tau_{m}} \pR + \Phi, \label{bound}
\end{equation}
from which we can derive the general relation between matching time $\tm$ and arbitrary constant $\Phi$:
\begin{equation}
\Phi = i\frac{Q}{\kappa}\ln \pL \kappa^{2}\tau_{m} \pR
\end{equation}
Here we have made use of the fact that $\ta = \ts - ia/\kappa$ and defined the imaginary time difference between $\ts$ and $\tm$ as $i\tau_{m} = \ts - \tm$.

If we take $\tm = \tK$ ($\tau_{m} = 1/\kappa^{2}$), as was derived in \cite{kaushal2013}, we get $\Phi = 0$, and
\begin{equation}
S_{C}^{\lr}\pL \rs \pR + G_{C}\pL \bp,T;\rs,\ta \pR = \int_{\tK}^{T}d\tau\,U\pL \int_{\ts}^{\tau}d\xi\,\bv(\ts) \pR. \label{eqn:phaseM1}
\end{equation}
However, with $\tm = \tQ$ ($\tau_{m} = Q/\kappa^{3}$), Eq.~\eqref{phSU} will lead to
\begin{equation}
S_{C}^{\lr}\pL \rs \pR + G_{C}\pL \bp,T;\rs,\ta \pR = \int_{\tQ}^{T}d\tau\,U\pL \int_{\ts}^{\tau}d\xi\,\bv(\ts) \pR + i\frac{Q}{\kappa}\ln\pL \frac{Q}{\kappa} \pR. \label{eqn:phaseM2}
\end{equation}
Equation~\eqref{eqn:phaseM1} and \eqref{eqn:phaseM2} are equivalent. In both cases, the long-range contribution of the asymptotic, Coulomb action $S_{C}^{\lr}$, matches with the long-range part of the quasiclassical eikonal-Volkov phase $G_{C}$, to give the final result independent of the mathematical 
construct of the $R$-matrix sphere radius. 
In the latter case, the second term will end up 
in the prefactor in the form $(Q/\kappa)^{Q/\kappa}$ (note that the co-ordinate $r_{Q}$ at time instant $\tQ$ is $Q/\kappa^{2}$), which is just the long-range prefactor term $(\kappa r)^{Q/\kappa}$ at coordinate $r_{Q} = Q/\kappa^{2}$. In the short-range limit, $Q \to 0$, this prefactor converges to unity.

Similar scheme for matching the Coulomb phase, used in the PPT method \cite{ppt1966,ppt1967,pp1967,pkp1968}, is derived for the quasistatic domain $\omega \to 0$ -- and subsequent approaches \cite{popru2008ii,murray2010}, the former using the imaginary time method (ITM) to derive ionization rates, the latter based on the partial Fourier transform scheme -- but the idea of introducing an arbitrary new constant was not considered in the way we have introduced here, especially the application to boundary matching for the momentum. See also recent review by Popruzhenko \cite{popru2014} for a comprehensive discussion of the plethora of theoretical approaches in Strong Field Ionisation, including study of Coulomb corrections to the short-range SFA and KFR theories.

\end{document}